\pdfoutput=1

\documentclass[10pt,fleqn]{article}\usepackage[]{graphicx}\usepackage[table, usenames, dvipsnames]{xcolor}
\makeatletter
\def\maxwidth{ %
  \ifdim\Gin@nat@width>\linewidth
    \linewidth
  \else
    \Gin@nat@width
  \fi
}
\makeatother

\definecolor{fgcolor}{rgb}{0.345, 0.345, 0.345}

\usepackage{framed}
\makeatletter
 {\par\unskip\endMakeFramed%
 \at@end@of@kframe}
\makeatother

\definecolor{shadecolor}{rgb}{.97, .97, .97}
\definecolor{messagecolor}{rgb}{0, 0, 0}
\definecolor{warningcolor}{rgb}{1, 0, 1}
\definecolor{errorcolor}{rgb}{1, 0, 0}
\newenvironment{knitrout}{}{} 

\usepackage{alltt}

\usepackage[margin=15pt,font=small,labelfont=bf]{caption}

\usepackage[T1]{fontenc}
\usepackage[utf8]{inputenc}
\usepackage{newpxtext,newpxmath}
\usepackage[scaled=0.85]{DejaVuSansMono}

\usepackage{amsmath}
\usepackage{booktabs}

\usepackage{subfig}

\usepackage[margin=25mm]{geometry}
\usepackage{setspace}

\usepackage{natbib}
\usepackage{lineno}

\usepackage[nolist]{acronym}

\usepackage{tikz}
\usetikzlibrary{shapes,arrows}

\DeclareGraphicsExtensions{.pdf,.PDF,.png,.PNG}

\DeclareMathOperator*{\argmax}{arg\,max}

\title{When does poor governance presage biosecurity risk?}

\usepackage{authblk}
\author[1]{Stephen E.\ Lane}
\author[2]{Tony Arthur}
\author[2]{Christina Aston}
\author[2]{Sam Zhao}
\author[1]{Andrew P.\ Robinson}
\affil[1]{Centre of Excellence for Biosecurity Risk Analysis, University of Melbourne, Parkville, Victoria 3010, Australia, \texttt{lane.s@unimelb.edu.au}}
\affil[2]{Department of Agriculture and Water Resources, Canberra, Australian Capital Territory 2601, Australia}

\usepackage[title]{appendix}

\usepackage{hyperref}
\definecolor{mylinkcolour}{HTML}{FC8D62}
\definecolor{myurlcolour}{HTML}{8DA0CB}
\definecolor{mycitecolour}{HTML}{66C2A5}
\hypersetup{
  linkcolor = mylinkcolour,
  urlcolor = myurlcolour,
  citecolor = mycitecolour,
  colorlinks = true
}
\IfFileExists{upquote.sty}{\usepackage{upquote}}{}
\begin{document}

\begin{acronym}
  \acro{AUC}{area under the curve}
  \acro{BRM}{biosecurity risk material}
  \acro{CSP}{continuous sampling plan}
  \acro{IPD}{inspections per detection}
  \acro{ROC}{Receiver Operating Characteristic}
\end{acronym}

\maketitle

\setlength\linenumbersep{10pt}
\renewcommand\linenumberfont{\normalfont\footnotesize\sffamily\color{MidnightBlue}}

\begin{abstract}

  Border inspection, and the challenge of deciding which of the tens of millions of consignments that arrive should be inspected, is a perennial problem for regulatory authorities. The objective of these inspections is to minimise the risk of contraband entering the country. As an example, for regulatory authorities in charge of biosecurity material, consignments of goods are classified before arrival according to their economic tariff number \citep{dibp2016}. This classification, perhaps along with other information, is used as a screening step to determine whether further biosecurity intervention, such as inspection, is necessary. Other information associated with consignments includes details such as the country of origin, supplier, and importer, for example.

  The choice of which consignments to inspect has typically been informed by historical records of intercepted material. Fortunately for regulators, interception is a rare event, however this sparsity undermines the utility of historical records for deciding which containers to inspect.

  In this paper we report on an analysis that uses more detailed information to inform inspection. Using quarantine biosecurity as a case study, we create statistical profiles using generalised linear mixed models and compare different model specifications with historical information alone, demonstrating the utility of a statistical modelling approach. We also demonstrate some graphical model summaries that provide managers with insight into pathway governance.
  
\end{abstract}

\section{Introduction}

Efficient and effective border biosecurity strategies are essential for protecting ecosystems and economies from invasive pests. The annual cost of invasive species generally is estimated to be over USD\$200bn \citep{Pimentel2011-pe} in the United States, and at least USD\$4bn in Australia \citep{Sinden_2005-yv}. In Australia, the Department of Agriculture and Water Resources (the department) is both the regulatory authority and the inspectorate for biosecurity protection, carrying out both pre-border and border intervention on a range of imported goods, based on the risk profile of the goods and international agreements. The objective of these interventions is to minimise the risk of \ac{BRM} entering the country.

Here, we focus on border inspection and the challenge of deciding which of the tens of millions of consignments that arrive should be inspected.  Before arrival, consignments of goods are classified according to their economic tariff number \citep{dibp2016}, and this classification is used, with other information, as a screening step to determine whether further biosecurity intervention, such as inspection, is necessary. Other information associated with consignments includes details such as the country of origin, supplier, and importer, for example.

Border inspection for quarantine biosecurity is carried out for a number of reasons, namely (i) to verify the effectiveness of mandated pre-arrival treatments; (ii) to detect and intercept \ac{BRM}; (iii) to provide information about the intrinsic contamination rate of the activity; and (iv) to deter potential malefactors.  As noted above, tens of millions of consignments arrive every year, so the challenge is to determine which should be inspected.

We define a \emph{pathway} as a collection of activities that culminates in the arrival to Australia of a set of alike consignments. The pathways are hierarchical, so we may consider a pathway of all consignments of a commodity, or all consignments of that commodity for a specific country, or even for a specific supplier. For example, the plant product pathway, which is the focus of this article, includes goods such as kiwi fruit and cashew nuts, which can themselves be considered pathways. Statistically, pathways can be thought of as processes.

Pathways can be classified as either high risk or a low risk, based on the probability that a consignment contains \ac{BRM}, called the \emph{approach rate}. For example, in Australia, kiwi fruit is a high-risk plant product pathway, with an approach rate of 55.8\%, whereas cashew nuts is a low-risk plant product pathway, with an approach rate of 1.3\%. Importantly, the degree of severity of the detected \ac{BRM} in cashews has been identified as very low.  The risk severity classification is important because the department may apply different interventions to low-risk than to high-risk pathways, as discussed below. The identification of pathways as high or low risk is called \emph{profiling}, and is an essential step in the efficient management of biosecurity intervention.

Traditionally, profiling has been applied by using records of interception of regulated pests on the pathway. This application is based on the assumption that future biosecurity compliance can be predicted by past biosecurity compliance, at least for some periods in the past and the future.  However, interception of regulated pests is a rare event, which is good news from the point of view of biosecurity protection, but makes profiling more difficult, especially in sparse pathways, because reliable estimates of pathway risk are hard to obtain.  This observation motivates the following question: whether future biosecurity compliance can be predicted by other characteristics as well as by past biosecurity compliance.

Historically, all consignments of imported plant product pathways were subjected to mandatory inspection. As part of a comprehensive review of Australia's biosecurity system, \citet{beale-2008} recommended establishing a science-based system for managing biosecurity issues, noting that zero risk is both unattainable and undesirable.  With the full inspection strategy, pathways that have lower approach rate cost considerably more inspection effort to intercept \ac{BRM}. For example, 4623 consignments were inspected in the cashew pathway over four years, of which \ac{BRM} was detected in 59, so the average number of \ac{IPD} was about 78, compared to about 2 for the kiwi fruit pathway.

We now introduce the case study that motivates the research. The inspection work flow of imported plant product pathways comprises three components, namely: suppliers that export plant products, importers that import the products from suppliers, and border inspections that attempt to detect as much as possible \ac{BRM}. Inspections at the border can be stratified by supplier or importer, that is, unique inspection regimes may be applied to individual importers or suppliers.

The department currently uses the \ac{CSP} algorithm, specifically, \ac{CSP}-3, to manage the biosecurity risk of low-risk pathways \citep{Dodge-1943, Dodge+Torrey-1951}. The \ac{CSP} family of algorithms allocate intervention effort within pathways according to recent inspection history. The department has implemented CSP-3 for the inspection of a range of low-risk pathways, including dried apricots, green coffee beans, raisins, cashews and some nuts. This particular approach to profiling has been shown to result in reductions of both leakage (how much \ac{BRM} is missed in the inspection process) and \ac{IPD} relative to random sampling plans \citep{acera1001j1, acera1101c1, acera1206f1}.

A wrinkle in the application of the CSP algorithm is that although it is implicit that the analysis of inspection history would take account of only the kinds of contamination that are of specific regulatory interest, in fact, any aspect of the inspection history can be used as an indicator of future risk.  That is, although the department may be specifically concerned about intercepting regulated pests, the inspection history provides a much richer view of the pathway because it includes information about other incidents, such as the interception of non-regulated pests, failures of documentation, and so on, which may arguably and testably be related to the chances of failure types that are of regulatory concern.  The question that motivated this study is: what data provide the most useful information about the pathway: the relatively sparse history of interception of regulated pests, or the more complete picture of the relative performance on the pathway, or some combination? Furthermore, can insight into the future performance in a given pathway be provided by information about historical performance in other, possibly related pathways?

This paper reports an analysis of the use of auxiliary information to try to improve profiling.  The objective is to distinguish high-risk and low-risk pathways, where risk refers to the interception rate of regulated pests, based on a range of characteristics of the pathway, including the interception rate of regulated pests, non-regulated pests, administrative failures, and supplier and tariff information.  We aim to form a picture of the governance of the pathway and use that picture as a basis for assessing their relative biosecurity risk.  The balance of the paper is organized as follows.  In the next section we introduce the dataset and the models used to test our conjectures.  We then present the results and a discussion and conclusion.

\section{Materials and methods}

We used a number of analytical approaches to assess the conjecture.  First, we tested the association between non-regulated pest and administrative (more frequent, low severity) failures and regulated pest inspection (less frequent, high severity) failures. If such an association was found, then we reasoned that historical governance-related variables may be used to predict future high-severity biosecurity failures. Second, we used historical failure rates to create profiles, and investigated performance using \ac{ROC} curves. Lastly, we constructed statistical models that would predict future regulated pest interception probabilities as a function of previous regulated pest interception probabilities and other, governance-related predictor variables.

All data preparation and modelling were performed using R Version 3.3.0 \citep{R_Core_Team2016-zb} with the generalised additive mixed models of Section~\ref{sec:modeling} using R package \texttt{rstanarm} \citep{rstanarm}.

\subsection{Data}\label{sec:data}

The data for the analysis comprise the inspection history for all consignments classified as fruit \citep[Chapter 8,][]{dibp2016} that arrived between January 2007 and December 2011, a period of five years.  The pathway is a complex one, comprising 80 different tariff codes, 3150 unique importers, and 3655 unique suppliers from 127 countries.  For the purposes of this study we will assume that all significant biosecurity contamination has been captured by the regulatory border inspection. There were approximately 48300 inspections of more than 75000 goods.  Approximately 5300 inspections resulted in interception of a regulated pest, 8500 inspections resulted in interception of a non-regulated pest, and 5900 inspections recorded some administrative failure.

For modelling (see Section \ref{sec:modeling}), we aggregated the data by year, tariff and supplier. This aggregation was done for two reasons: first, it allowed us to create models that account for both supplier and tariff effects, and second, aggregating by year limits the effects of any seasonality. We use \textit{interceptions} to refer to both interceptions of pests and administrative failures throughout the study. An appropriately formatted dataset for modelling was constructed as follows:

\noindent For each year $y$ within 2008 to 2011;
\begin{itemize}
\item Compute interception/fail rates for year $y - 1$ by tariff, supplier, and year for:
  \begin{itemize}
  \item Administrative interceptions
  \item Non-regulated pest interceptions
  \item Regulated pest interceptions
  \end{itemize}
\end{itemize}

We denote by $X_{sty}$ the number of interceptions out of $n_{sty}$ inspections from tariff $t$ performed in year $y$ from supplier $s$. Correspondingly, each inspection has a probability $p_{sty}$ of being intercepted in one of the ways listed above. Then $X_{sty}$ was modelled as
\begin{align*}
  X_{sty} & \stackrel{d}{=}  \mbox{Binomial}(p_{sty}, n_{sty})
\end{align*}

Computing interception rates by tariff, supplier and year sometimes resulted in very small binomial denominators, due to the sparse history of inspection and interceptions produced. For this reason, rather than \emph{raw} interception rates, we calculated \emph{smoothed} interception rates using parametric empirical Bayes \citep{Robinson2015-oa}. In particular, we used the Beta-binomial model to smooth interception rates for suppliers within tariffs and years; we provide the full details in Appendix \ref{sec:calc-rates-using}.

\subsection{Association between low-severity and high-severity interceptions}

To investigate the association between low-severity and high-severity interceptions, we calculated log-odds ratios and 95\% confidence intervals (using a normal approximation for the log-odds) for the odds of a regulated-pest interception for consignments with or without non-regulated pest or administrative interceptions.

For each inspected consignment, suppose $Y$ denotes the outcome of inspection, so that $Y=1$ indicates a regulated pest was intercepted. Further, let $X$ denote whether the consignment contained a non-regulated pest (or had an administrative failure), so that $X=1$ indicates the consignment contains a non-regulated pest (or had an administrative failure). The log-odds ratio is
\begin{align*}
  \log\text{OR} & = \log\left[\frac{\Pr(Y=1|X=1)/\Pr(Y=0|X=1)}{\Pr(Y=1|X=0)/\Pr(Y=0|X=0)}\right].
\end{align*}

\subsection{Profiling using annual inspection data}\label{sec:model-comp-polic}

We created profiles using annual inspection data, and compared performance using \ac{ROC} curves. For each year $y$ in 2007 to 2010 and each kind of interception rate (regulated pest, non-regulated pest and administrative) we: compute \ac{ROC} curves against year $y+1$ biosecurity inspection outcomes for regulated pests and further, calculate the \ac{AUC}.

We also computed \ac{ROC} curves within tariff, due to the suspicion that the tariff-to-tariff variation would dominate the \ac{ROC} signal, due to the differences of interception rates between the tariffs, rather than for the importers within the tariffs. That is, if we only ran the profiles across the tariffs then a naive assessment of the performance would look very good because we would expect the differences between the risks of the tariffs to be reasonably stable from year to year. Hence, assessing the model within tariffs provides a more reasonable assessment.

\subsection{Profiling using statistical modelling}\label{sec:modeling}

We chose to construct models using generalised additive mixed model formulations with the linear predictors for the logit probability, $\log\left(\frac{p_{sty}}{1 - p_{sty}} \right)$ specified as:
\begin{align*}
  \text{Base} & : \beta_0 + \gamma_s + \tau_t \\
  \text{M1} & : \beta_0 + \gamma_s + \tau_t + \alpha_{y} \\
  \text{M2} & : \beta_0 + \gamma_s + \tau_t + \alpha_{y} + \varphi_{st} \\
  \text{M3} & : \beta_0 + \gamma_s + \tau_t + \alpha_{y} + \varphi_{st} + \kappa_{sy} \\
  \text{M4} & : \beta_0 + \gamma_s + \tau_t + \alpha_{y} + \varphi_{st} + \kappa_{sy} + b_3 (p_{R, st(y - 1)}) \\
  \text{M5} & : \beta_0 + \gamma_s + \tau_t + \alpha_{y} + \varphi_{st} + \kappa_{sy} + b_3 (p_{N, st(y - 1)}) \\
  \text{M6} & : \beta_0 + \gamma_s + \tau_t + \alpha_{y} + \varphi_{st} + \kappa_{sy} + b_3 (p_{A, st(y - 1)})
\end{align*}

\noindent where:
$\beta_0$ is a fixed process constant to be estimated;
$\gamma_s$ is a supplier-level effect;
$\tau_t$ is a tariff-level effect;
$\alpha_y$ is a effect for year of interception;
$\varphi_{st}$ is a effect for the supplier-tariff cross-classification;
$\kappa_{sy}$ is a effect for the supplier-year cross-classification;
$b_3(\cdot)$ represent cubic regression splines for the previous year's regulated pest interception rate $p_{R, st(y - 1)}$, non-regulated pest interception rate $p_{N, st(y - 1)}$, and administrative interception rate $p_{A, st(y - 1)}$. Bayesian logistic regression models were fit using \texttt{rstanarm} \citep{rstanarm}. We used student-$t$ priors for all coefficients, setting the scale for the intercept prior at 10, and for all other coefficients at 2.5.

To be more descriptive, (M4) tests whether historical regulated pest interception rates can be used to predict future regulated pest interception probability, whilst (M5) and (M6) test effect of historical non-regulated pest and administrative interception rates on probability of future regulated pest interception.

Comparison of the statistical profiling results was made via a combination of: LOOIC comparisons \citep{Vehtari2016-dp}, and predictive log-likelihood via repeated five-fold cross-validation. LOOIC is similar to AIC in that it estimates out-of-sample prediction accuracy; however, LOOIC integrates over uncertainty in the parameters, and does not assume multivariate normality as the AIC does. We used 20 repeats, resulting in 100 training/testing datasets for comparison. To ensure balance across the datasets, sampling was performed within years. All models from Section~\ref{sec:modeling} were fit to each training dataset, and predictions made on the testing datasets.

\section{Results}

We present the results in three sections: the association between low-severity and high-severity interceptions; the operational \ac{AUC} tests and the statistical modelling results. We finish this section with an in-depth look at the information gained from the modelling procedures.

\subsection{Association between low-severity and high-severity interceptions}
\label{sec:assoc-betw-low}

\begin{figure}[!h]
  \centering
  \includegraphics[width=160mm]{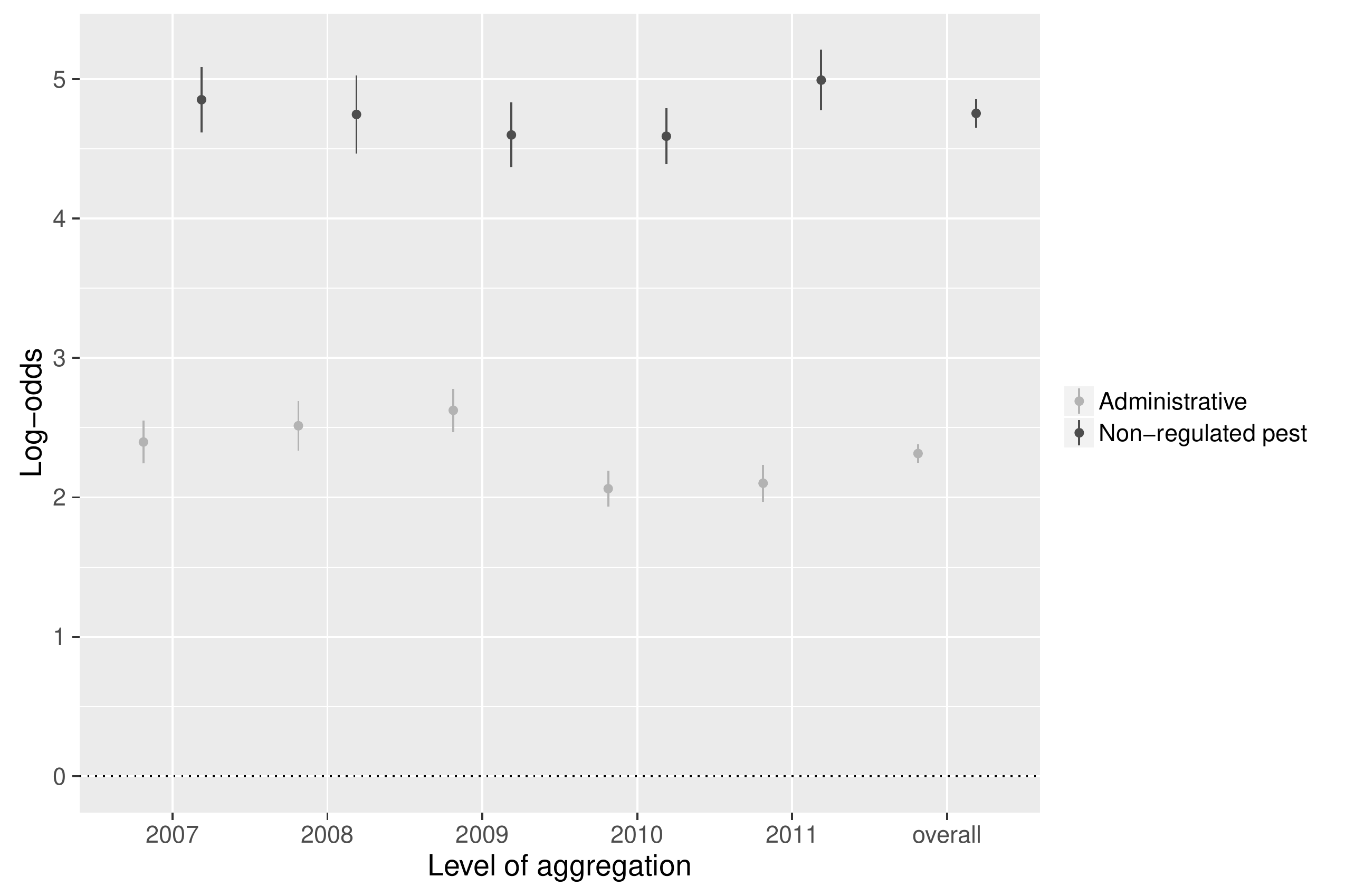}
  \caption[Associations between low- and high-risk inspection failures.]{Estimates (95\% confidence intervals) of the log-odds ratios between low- and high-risk interceptions overall and by year (2007--2011). The log-odds ratios are calculated between regulated pest (high-risk) interceptions, and non-regulated pest and administrative (low-risk) interceptions.}
  \label{fig:odds}
\end{figure}

Figure~\ref{fig:odds} shows the log-odds ratios, along with 95\% confidence intervals for the association between regulated pest (high-risk) interceptions and non-regulated pest and administrative (low-risk) interceptions both overall and by year. All estimates and lower bounds of the confidence intervals are well above 0, showing there is a large association between low- and high-risk interceptions.

\subsection{Comparison of profiles using annual inspection data}
\label{sec:oper-sign}


\subsubsection{Across tariffs}

\begin{figure}[!h]
  \centering
  \includegraphics[width=160mm]{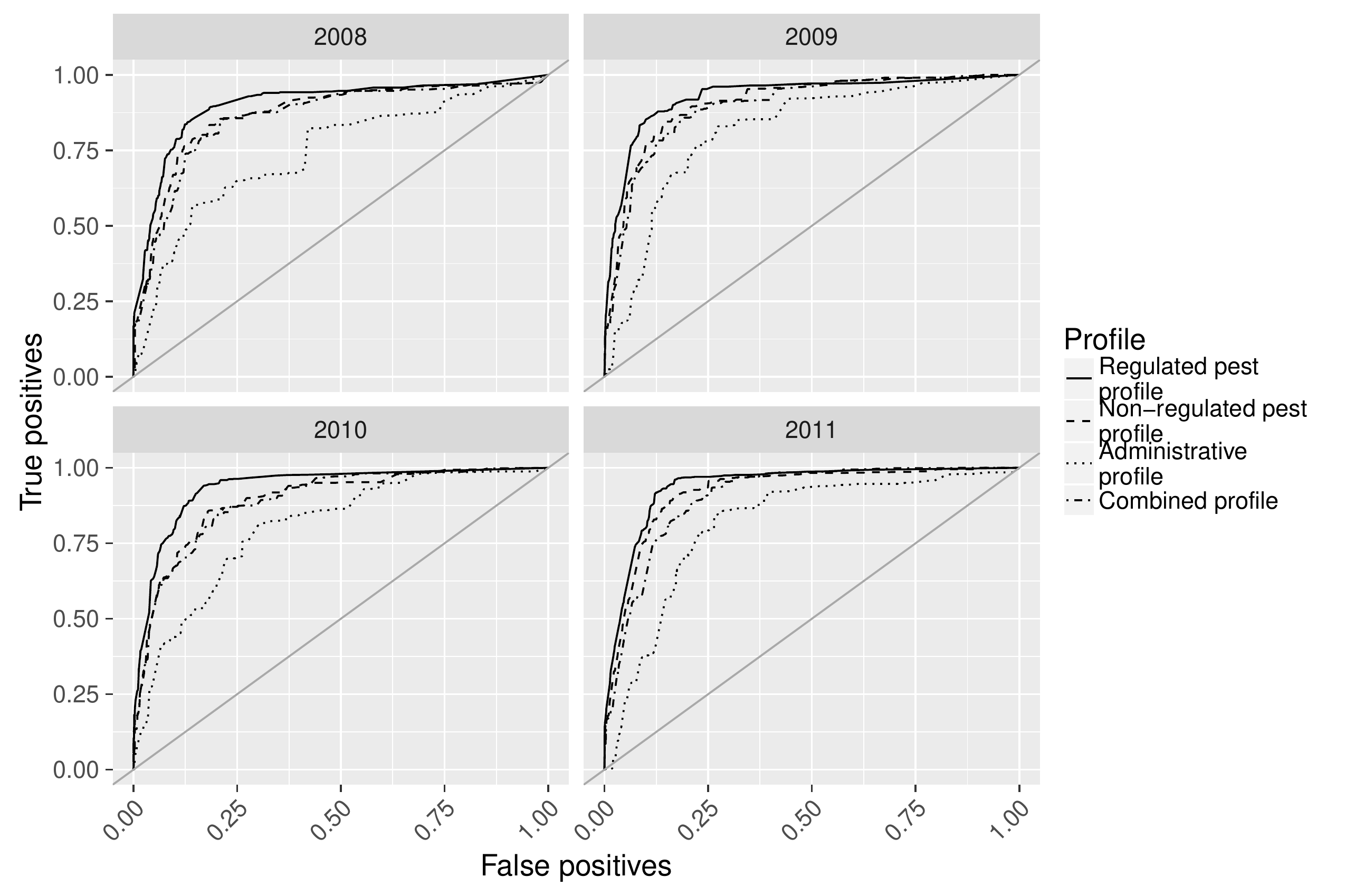}
  \caption[ROC curves showing the performance of four profiling strategies for four years of data (2007--2011).]{ROC curves showing the performance of four profiling strategies for four years of data (2007--2011). The profiles are constructed by tariff and importer using the previous year's inspection data. A line is added at $x = y$ to facilitate comparison.}
  \label{fig:roc}
\end{figure}

Figure~\ref{fig:roc} presents ROC curves that compare how well the different profiles perform. As per Section~\ref{sec:model-comp-polic}, the profiles are generated from the previous year's interception rates. All profiling approaches are substantially better than random, and the administrative profile consistently led to the weakest performance across each year. We have also shown the performance from a \emph{combined} profile in Figure~\ref{fig:roc}; this is simply the profile using interception rates calculated from a variable indicating if \emph{any} of the interception types occur. Clearly, the combined interception profile offers little performance over the regulated-pest profile.

The profiles derived from non-regulated pest and administrative interception rates were consistently slightly worse than those based on regulated pest interception rates. Table~\ref{tab:aucsEB} presents the \ac{AUC} values for each of the curves presented in Figure~\ref{fig:roc}. The values are consistently close to 1, which suggests that the relative interception rates are very stable from year to year, and that the interception rates for each year $y$ are a very good indicator for year $y+1$. We also derived profiles using data without empirical Bayes smoothing (see Section~\ref{sec:data}), however these profiles underperformed compared to the profiles using empirical Bayes smoothing. The results without empirical Bayes smoothing are shown in Appendix~\ref{sec:further-results}, Table~\ref{tab:aucsAll}.

\begin{table}[!h]
\caption{Summary of AUC values for profiling strategies, by year.  The profiles are as follows: \emph{Regulated pest} refers to using the previous year's regulated pest interception rate; \emph{Non-regulated pest} refers to using the previous year's non-regulated pest interception rate; \emph{Administrative} refers to using the previous year's administrative interception rate; and \emph{Combined} refers to using the previous year's combined interception rate. Each AUC is computed using the data from the \emph{following} year's inspections.\label{tab:aucsEB}} 
\begin{center}
\begin{tabular}{lrrrr}
\toprule
\multicolumn{1}{l}{Profile}&\multicolumn{1}{c}{2008}&\multicolumn{1}{c}{2009}&\multicolumn{1}{c}{2010}&\multicolumn{1}{c}{2011}\tabularnewline
\midrule
Regulated pest&0.902&0.929&0.935&0.939\tabularnewline
Non-regulated pest&0.870&0.909&0.892&0.920\tabularnewline
Administrative&0.743&0.815&0.803&0.813\tabularnewline
Combined&0.859&0.896&0.890&0.905\tabularnewline
\bottomrule
\end{tabular}\end{center}
\end{table}


\subsubsection{Within tariffs}

\begin{figure}[!h]
  \centering
  \includegraphics[width=150mm]{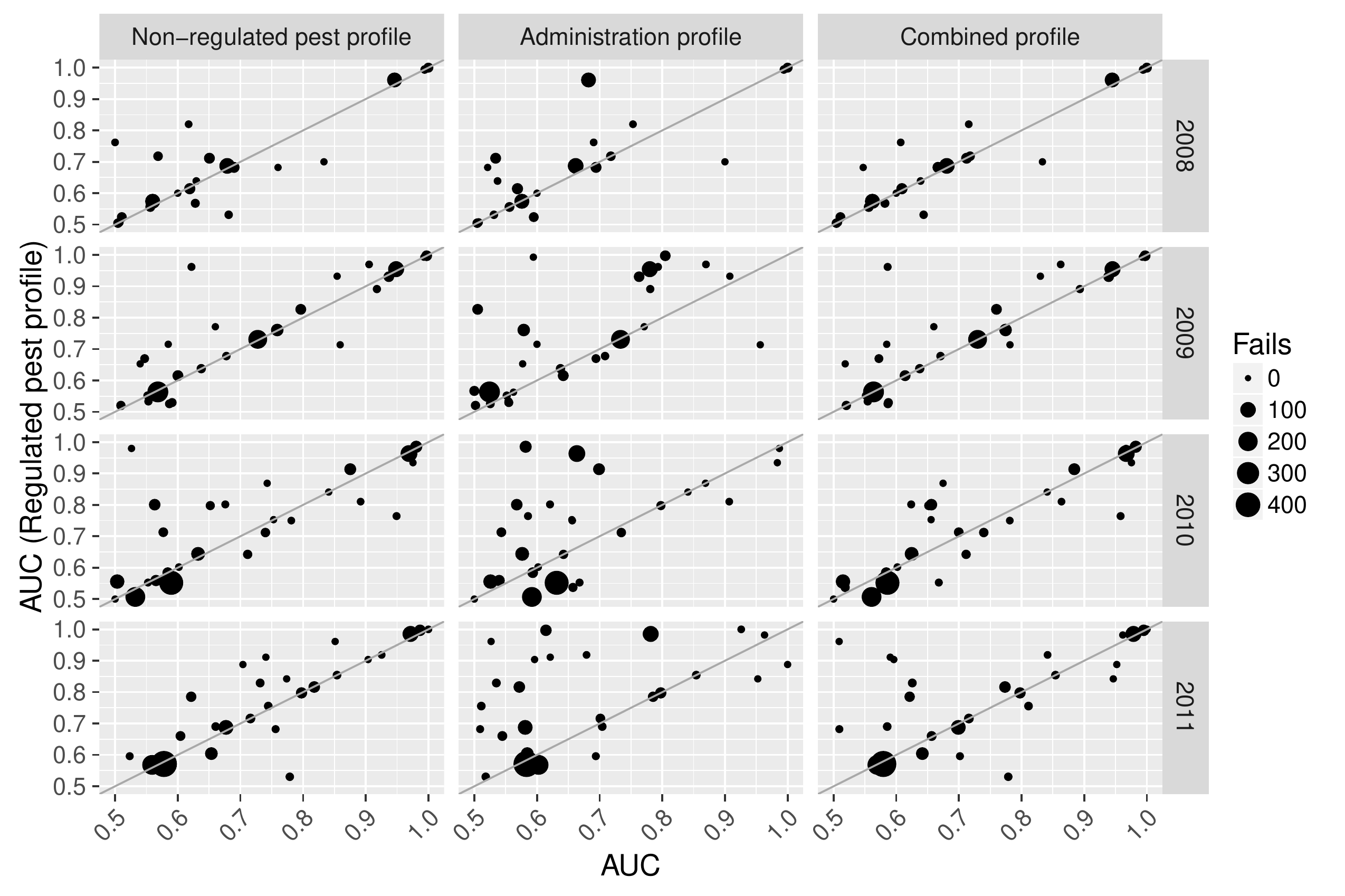}
  \caption{\ac{AUC}s computed for each tariff code in the data to assess within-tariff profiling operationally, by year. The $y$-axis is the \ac{AUC} using the previous year's regulated pest interception rate to set the profile. The $x$-axis in each panel is the \ac{AUC} using the previous year's non-regulated pest, administration, and combined interception rates to set the profile, assessed on the same inspection data that are used for the $y$-axis. The size of the point is related to the number of fails within the profile, and a line has been added at $x = y$ to facilitate comparison.}
  \label{fig:auc-eb}
\end{figure}

As noted in Section~\ref{sec:model-comp-polic}, our suspicion was that tariff-to-tariff variation would dominate the \ac{ROC} signal, evidence for which was supported via modelling (Table~\ref{tab:anova}). Figure~\ref{fig:auc-eb} plots the \ac{AUC}s arising from the regulated pest profile vs. the \ac{AUC}s arising from the non-regulated pest, administrative and combined profiles respectively, within tariff. Each point represents an \ac{ROC} curve applied to a single tariff, where the entities within the tariff that are being profiled are the suppliers. The size of each point indicates the number of regulated pest interceptions in the tariff, providing a sense of importance of that tariff.

A relationship between the number of regulated pest interceptions and AUC is not apparent in Figure~\ref{fig:auc-eb}; we would expect larger points in the top right corner if this were the case. However, within-tariff variation is considerable, providing a measure of conservatism against the strong performance of the across-tariff comparisons shown in Table~\ref{tab:aucsEB}. This suggests that any profiling undertaken would need to take account of the tariff being profiled.

Correlation between the regulated pest profile and the other profiling strategies appears strong, especially between the non-regulated pest and combined profiles. The administrative profile results however, show that many of the regulated pest \ac{AUC}s lie above the $y=x$ line. This suggests that the administrative profiles are likely to perform worse than the non-regulated pest profiles. This observation is supported by the findings of the statistical modelling (Table~\ref{tab:anova}), where the model that included non-regulated pest interception rates performed better than the model including administrative interception rates.

\subsection{Comparison of profiles using statistical modelling}
\label{sec:model-based-sign}

Comparison of the models from Section~\ref{sec:modeling} is reported in Table~\ref{tab:anova}. Model M3 has the lowest LOOIC, and the difference in LOOIC between M3 and M4 (98.3) is much larger than the standard error of its difference (17). These results show that supplier and tariff information are important for predicting regulated pest interception probability. The models are greatly improved with the addition of interaction terms between suppliers and tariffs, and suppliers and years. After allowing for the effects of suppliers, tariffs and years, the addition of: the previous year's regulated pest interception rate (M4 vs M3), the previous year's non-regulated pest interception rate (M5 vs M3) and the previous year's administrative interception rates (M6 vs M3) do not improve the model.

\begin{table}[!h]
\caption[LOOIC-based comparison of GAMM models.]{LOOIC-based comparison of statistical profiling models. The model with the smallest LOOIC (M3) is shown first, with subsequent rows ordered by increasing LOOIC. $\Delta$LOOIC shows the difference in LOOIC between all models and model M3; se($\Delta$LOOIC) shows the estimated standard error of the difference. Eff. P gives the estimated effective number of parameters; se(Eff. P) shows its standard error.\label{tab:anova}} 
\begin{center}
\begin{tabular}{lrrrrrr}
\toprule
\multicolumn{1}{l}{Model}&\multicolumn{1}{c}{LOOIC}&\multicolumn{1}{c}{se(LOOIC)}&\multicolumn{1}{c}{Eff. P}&\multicolumn{1}{c}{se(Eff. P)}&\multicolumn{1}{c}{$\Delta$LOOIC}&\multicolumn{1}{c}{se($\Delta$LOOIC)}\tabularnewline
\midrule
M3&$2788$&$111$&$458$&$23.3$&$$&$$\tabularnewline
M4&$2886$&$117$&$470$&$25.0$&$ 98.3$&$17.0$\tabularnewline
M2&$2980$&$127$&$419$&$23.6$&$192.2$&$29.4$\tabularnewline
M1&$3093$&$133$&$377$&$22.5$&$304.9$&$36.6$\tabularnewline
M5&$3141$&$146$&$449$&$26.6$&$353.7$&$49.6$\tabularnewline
M6&$3154$&$145$&$446$&$26.9$&$366.2$&$49.6$\tabularnewline
Base&$3250$&$148$&$392$&$23.7$&$462.7$&$52.3$\tabularnewline
\bottomrule
\end{tabular}\end{center}
\end{table}

Figure~\ref{fig:mod-aucs} shows the out-of-sample mean log predictive density and AUCs (Section~\ref{sec:model-comp-polic}) for all statistical profiling methods. Also shown in Figure~\ref{fig:mod-aucs2} are the AUCs from the regulated pest profile. Models Base--M3 perform the best in terms predictive log-likelihood (larger values are better), with no clear demarcation between them. In comparison, Models M1 and M2, as well as the Base model and the empirical Bayes profile perform best on AUC.

\begin{knitrout}
\definecolor{shadecolor}{rgb}{0.969, 0.969, 0.969}\color{fgcolor}\begin{figure}

{\centering \subfloat[Mean log predictive density\label{fig:mod-aucs1}]{\includegraphics[width=0.45\textwidth]{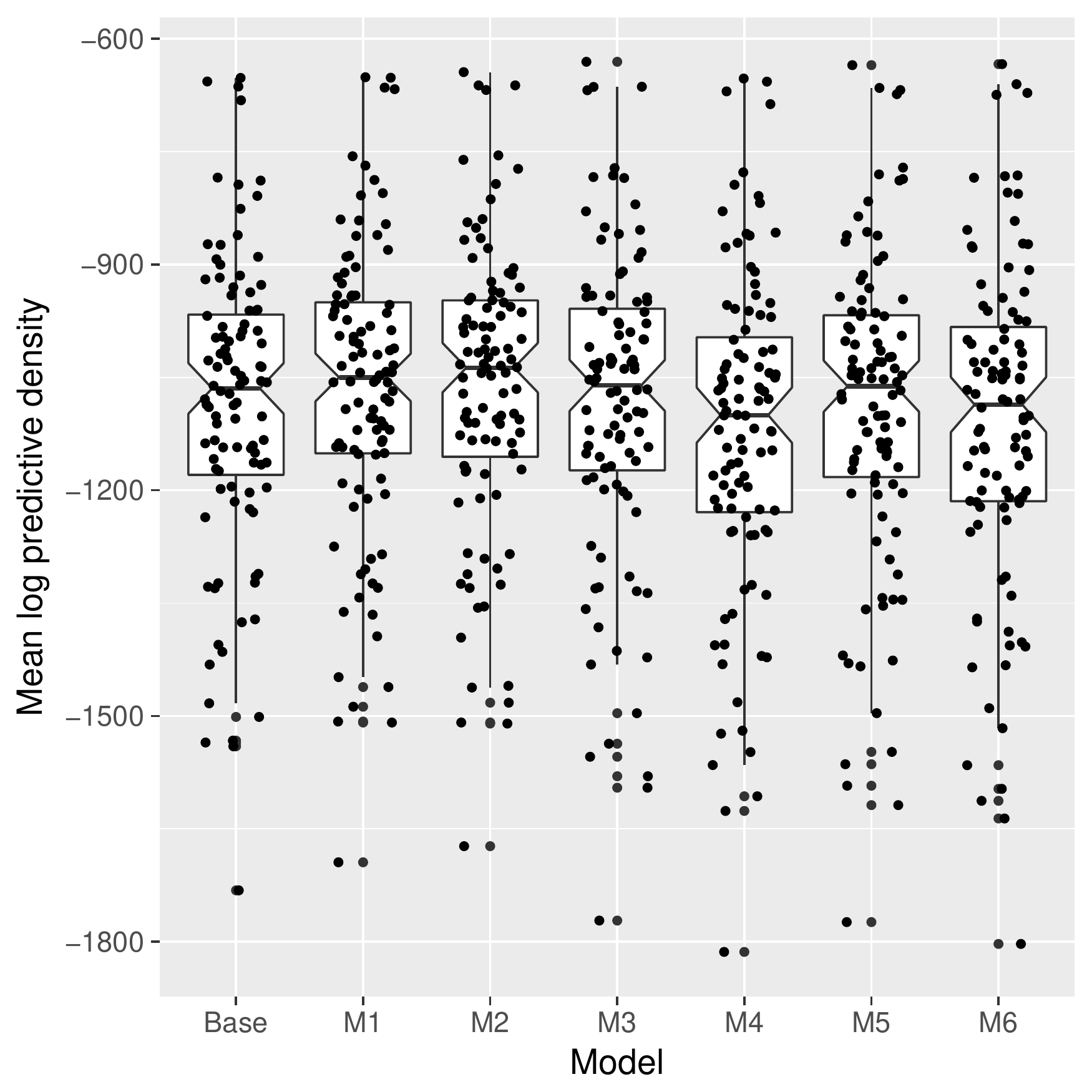} }
\subfloat[AUCs\label{fig:mod-aucs2}]{\includegraphics[width=0.45\textwidth]{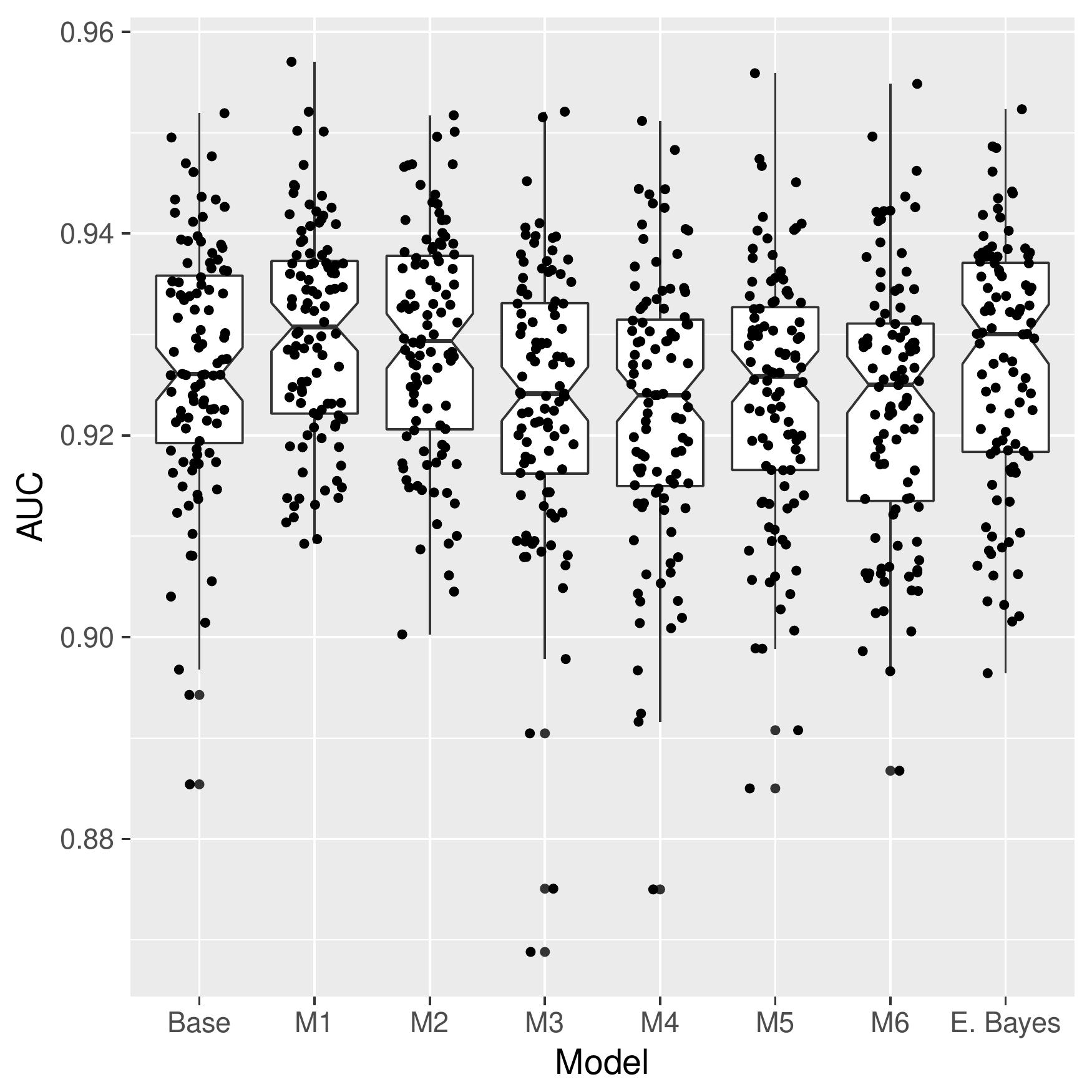} }

}

\caption[Mean log predictive density and AUCs for statistical profiling]{Mean log predictive density and AUCs for statistical profiling. Also shown are the AUCs from the regulated pest profile (E. Bayes).}\label{fig:mod-aucs}
\end{figure}

\end{knitrout}

\subsection{Model examination}
\label{sec:model-examination}

In this section we present an investigation of the effects from Model M3. We decided to investigate Model 3 further due to its superior performance in LOOIC (Table~\ref{tab:anova}), as well as the within-supplier examinations that would be available due to the interaction term. Figure~\ref{fig:tariffs} shows the marginal log-odds for tariffs from Model M3, ordered left-to-right by decreasing probability of their marginal log-odds being greater than 0; bars in the figure show 90\% posterior credible intervals. The inset shows the top 10 tariffs, and as to be expected, Kiwi fruit is the tariff that contributes the highest risk.


\begin{knitrout}
\definecolor{shadecolor}{rgb}{0.969, 0.969, 0.969}\color{fgcolor}\begin{figure}[!hbp]

{\centering \includegraphics[width=\maxwidth]{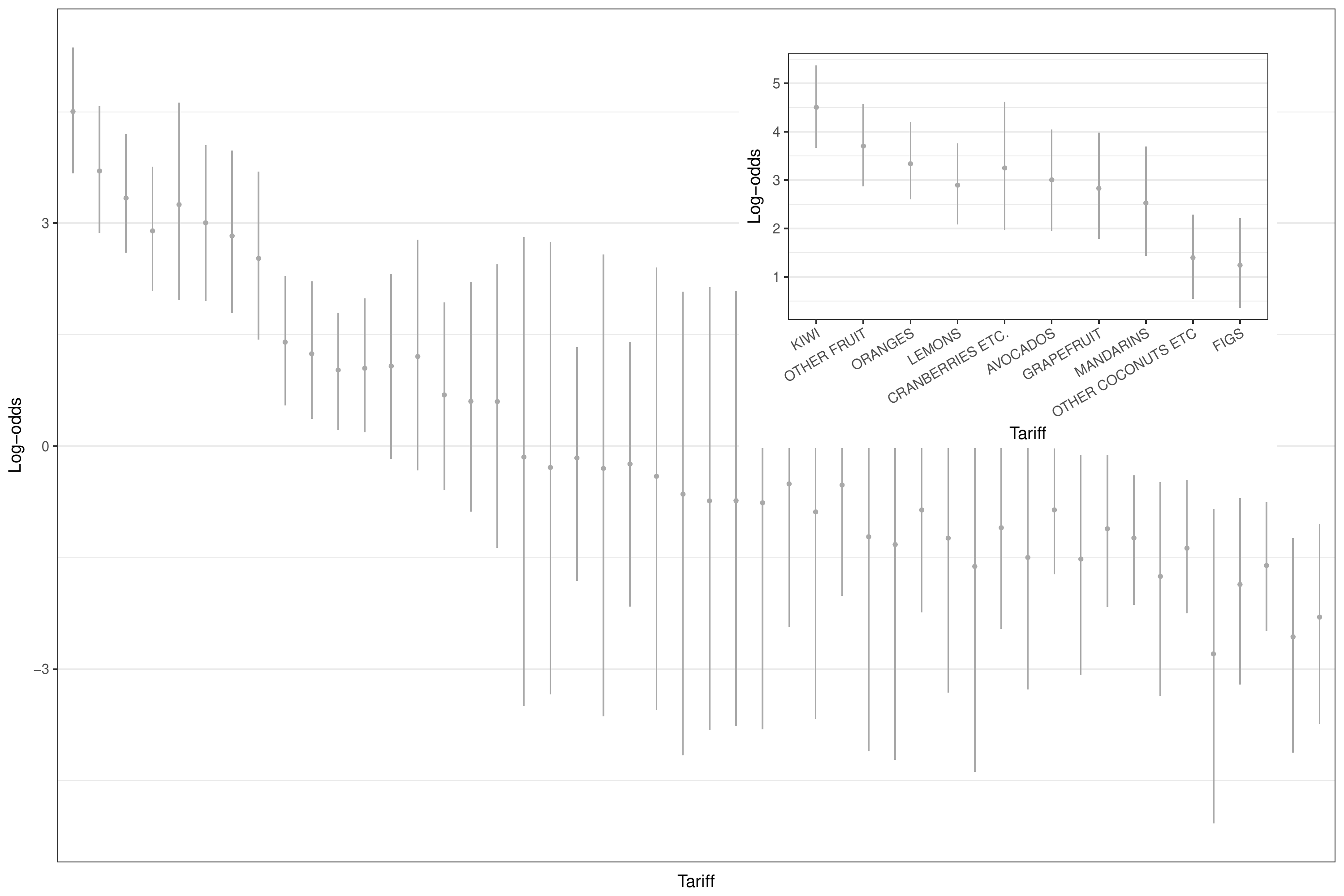} 

}

\caption[Marginal log-odds of the tariff effect in Model 3]{Marginal log-odds of the tariff effect in Model 3. Tariffs are ordered left-to-right by decreasing probability of their marginal log-odds being greater than 0, with bars showing 90\% posterior credible intervals. The inset shows the top 10 tariffs.}\label{fig:tariffs}
\end{figure}

\end{knitrout}

Figure~\ref{fig:suppliers} shows the marginal log-odds for suppliers from the model, ordered left-to-right by decreasing posterior probability of their marginal log-odds being greater than 0; bars in the figure show 90\% posterior credible intervals. Supplier labels have been masked for privacy reasons. The suppliers to the left of the figure are those predicted to have a large increase in probability of regulated pest interception, all else being equal. It is these suppliers that would naturally be the first targets in an operational capacity.

\begin{knitrout}
\definecolor{shadecolor}{rgb}{0.969, 0.969, 0.969}\color{fgcolor}\begin{figure}
\includegraphics[width=\maxwidth]{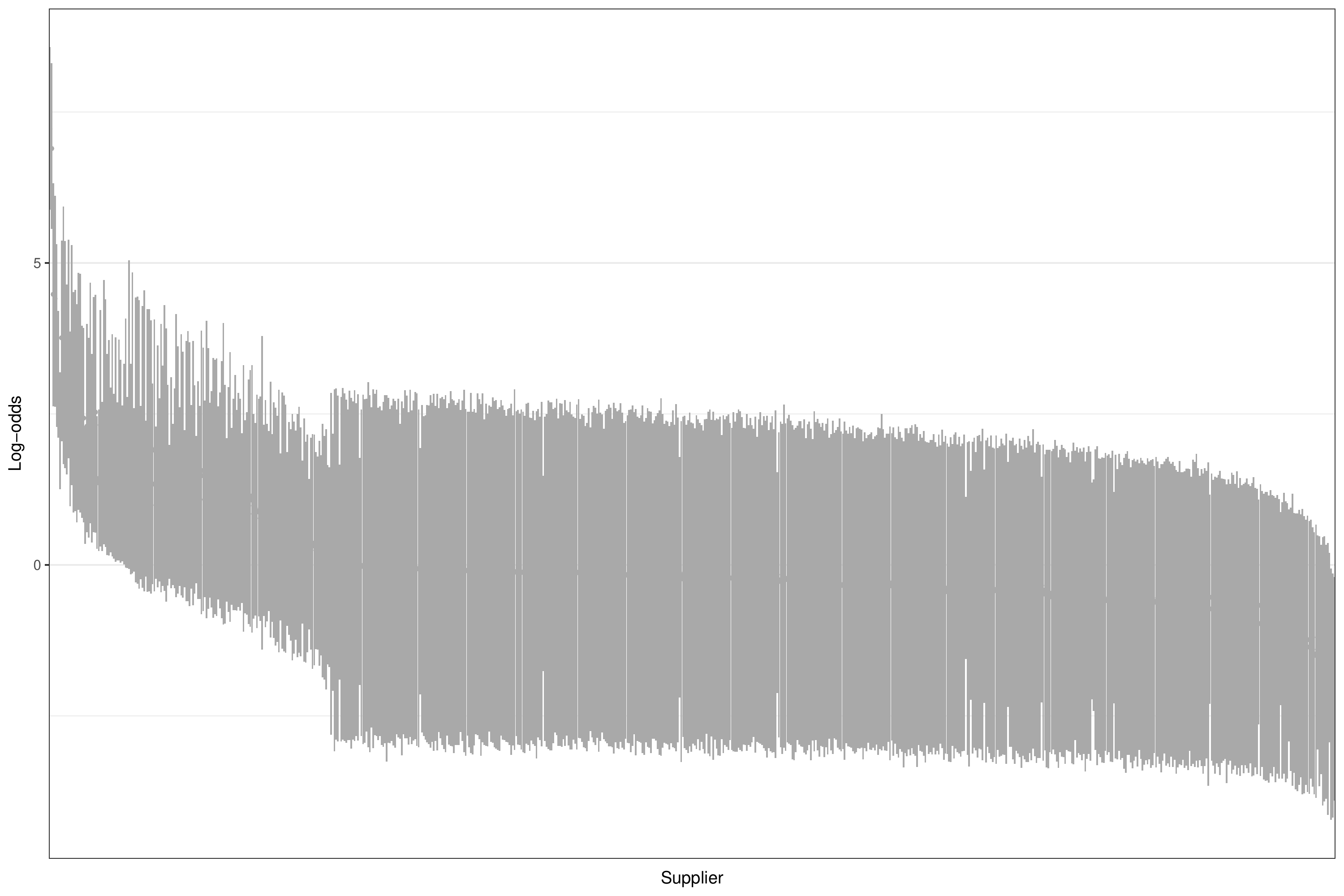} \caption[Marginal log-odds of the supplier effect in Model M3]{Marginal log-odds of the supplier effect in Model M3. Suppliers are ordered left-to-right by decreasing posterior probability of their marginal log-odds being greater than 0, with bars showing 90\% posterior credible intervals. The inset shows the top 10 tariffs.}\label{fig:suppliers}
\end{figure}

\end{knitrout}

Figure~\ref{fig:suppTarfs} provides a closer examination of the risky suppliers. We have selected the top 25 suppliers (by the probability of their marginal log-odds being greater than 0) and calculated their posterior probability of a regulated pest being present in a consignment, averaged over all years from Model M3. The panel on the left shows their posterior probability for each tariff that the supplier imports, whilst the panel on the right shows the observed proportion (averaged over years) of regulated pest interceptions by tariff. This figure shows that these highest risk suppliers import a range of tariffs --- i.e. their poor performance is not necessarily due to importing one or two of the highest risk tariffs (as shown in Figure~\ref{fig:tariffs}). Further, in the tariffs they do import, they have consistently high levels of consignments with regulated pest contamination (right panel, Figure~\ref{fig:suppTarfs}).

\begin{knitrout}
\definecolor{shadecolor}{rgb}{0.969, 0.969, 0.969}\color{fgcolor}\begin{figure}
\subfloat[Posterior probability of a regulated pest interception.\label{fig:suppTarfs1}]{\includegraphics[width=0.49\linewidth]{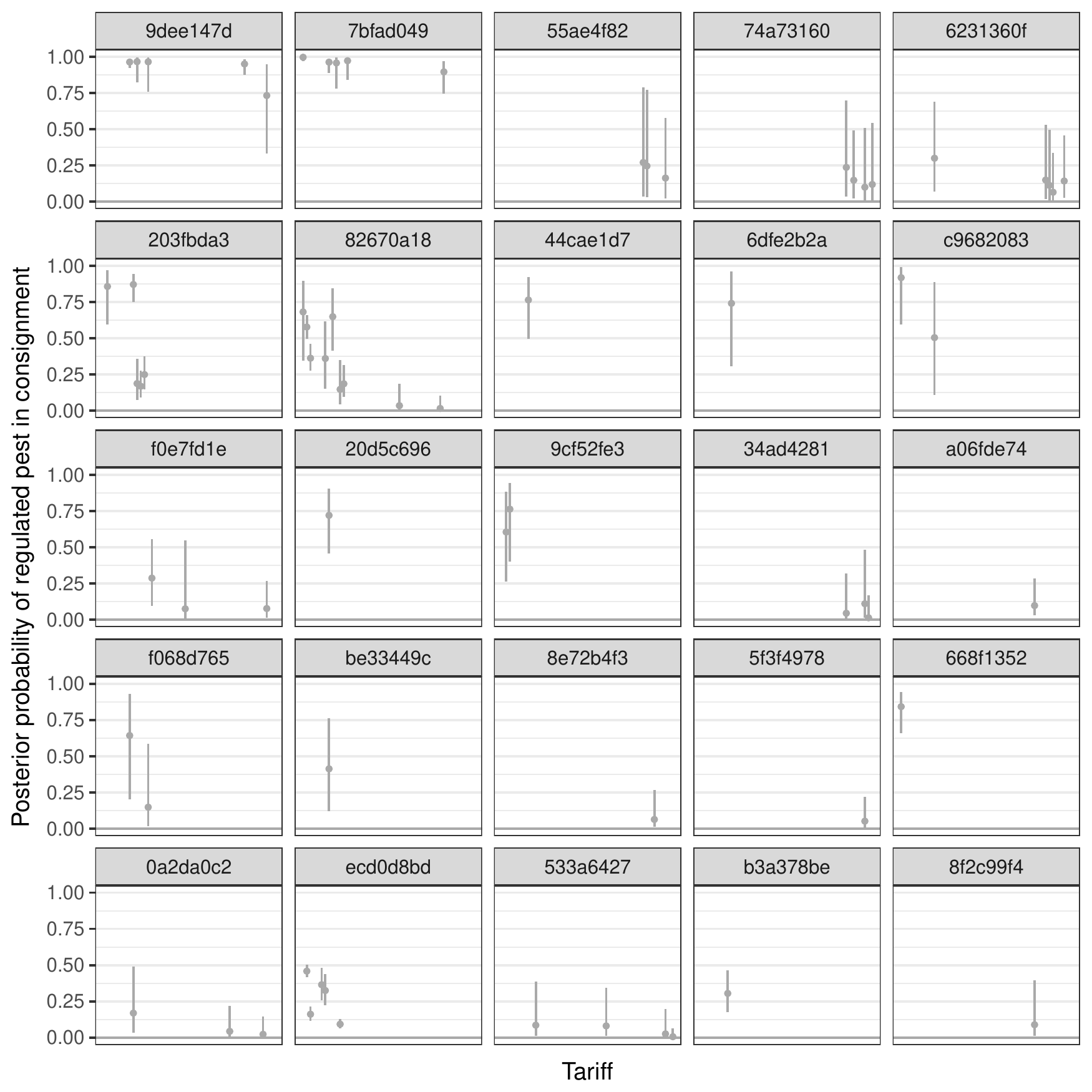} }
\subfloat[Observed proportion of consignments with regulated pests.\label{fig:suppTarfs2}]{\includegraphics[width=0.49\linewidth]{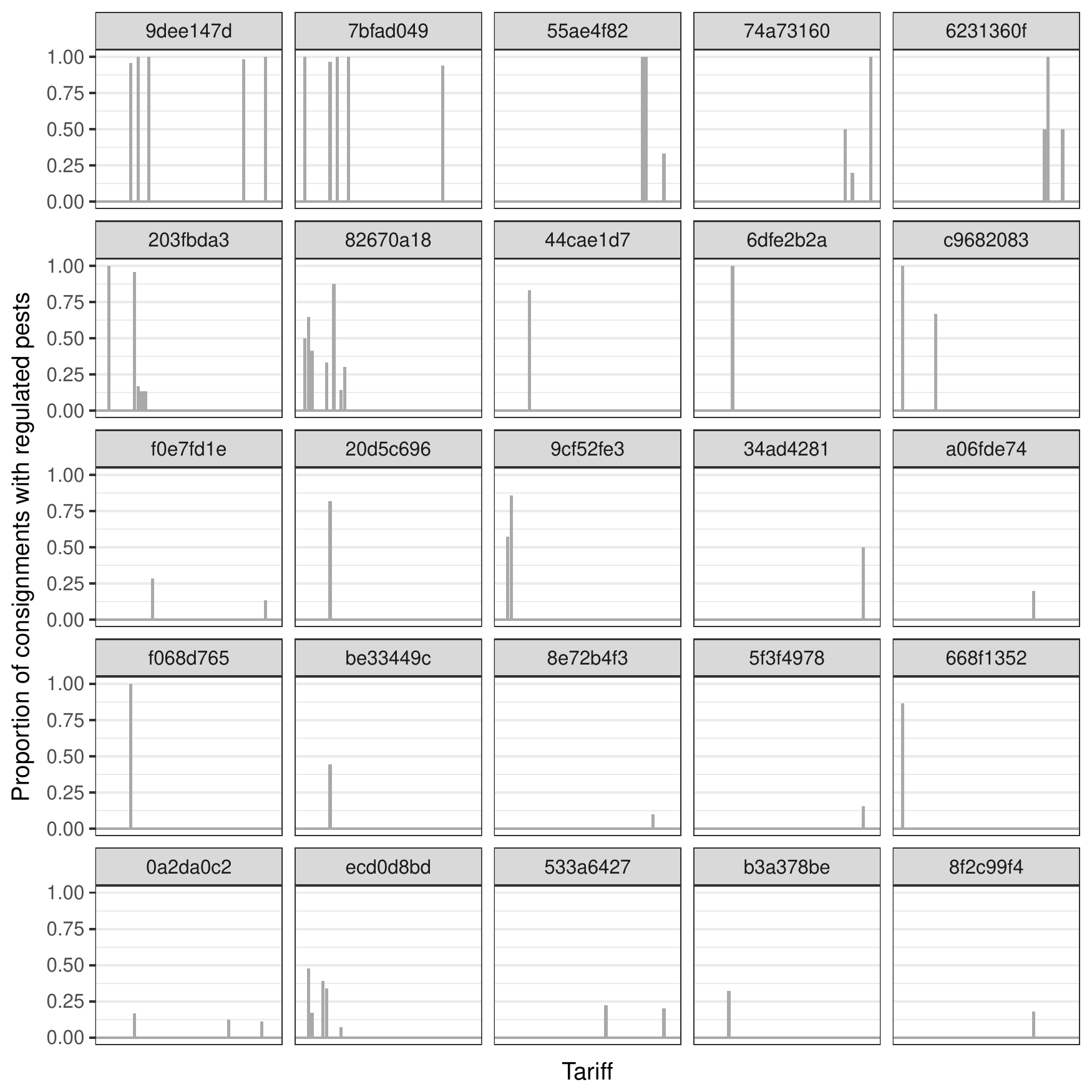} }\caption[Posterior probability of a regulated pest interception in the top 25 suppliers, along with the observed proportion of consignments with regulated pests]{Posterior probability of a regulated pest interception in the top 25 suppliers, along with the observed proportion of consignments with regulated pests. Panels are ordered left-to-right, top-to-bottom by the probability of their marginal log-odds being greater than 0; bars in the left panel show 90\% posterior credible intervals.}\label{fig:suppTarfs}
\end{figure}

\end{knitrout}

Figure~\ref{fig:suppTarfsBot} provides a closer examination of suppliers who pose minimal risk. We have selected the bottom 25 suppliers (by the probability of their marginal log-odds being greater than 0) and calculated their posterior probability of a regulated pest being present in a consignment, averaged over all years from Model M3. The panel on the left shows their posterior probability for each tariff that the supplier imports, whilst the panel on the right shows the observed proportion (averaged over years) of consignments that \textit{did not} contain regulated pests by tariff. Similar to Figure~\ref{fig:suppTarfs}, these suppliers import a range of tariffs --- i.e. their good performance is not necessarily due to importing lower risk tariffs. However, in comparison to the risky suppliers, in the tariffs they do import, they have consistently high levels of consignments without regulated pest contamination (right panel, Figure~\ref{fig:suppTarfsBot}).

\begin{knitrout}
\definecolor{shadecolor}{rgb}{0.969, 0.969, 0.969}\color{fgcolor}\begin{figure}
\subfloat[Posterior probability of a regulated pest interception.\label{fig:suppTarfsBot1}]{\includegraphics[width=0.49\linewidth]{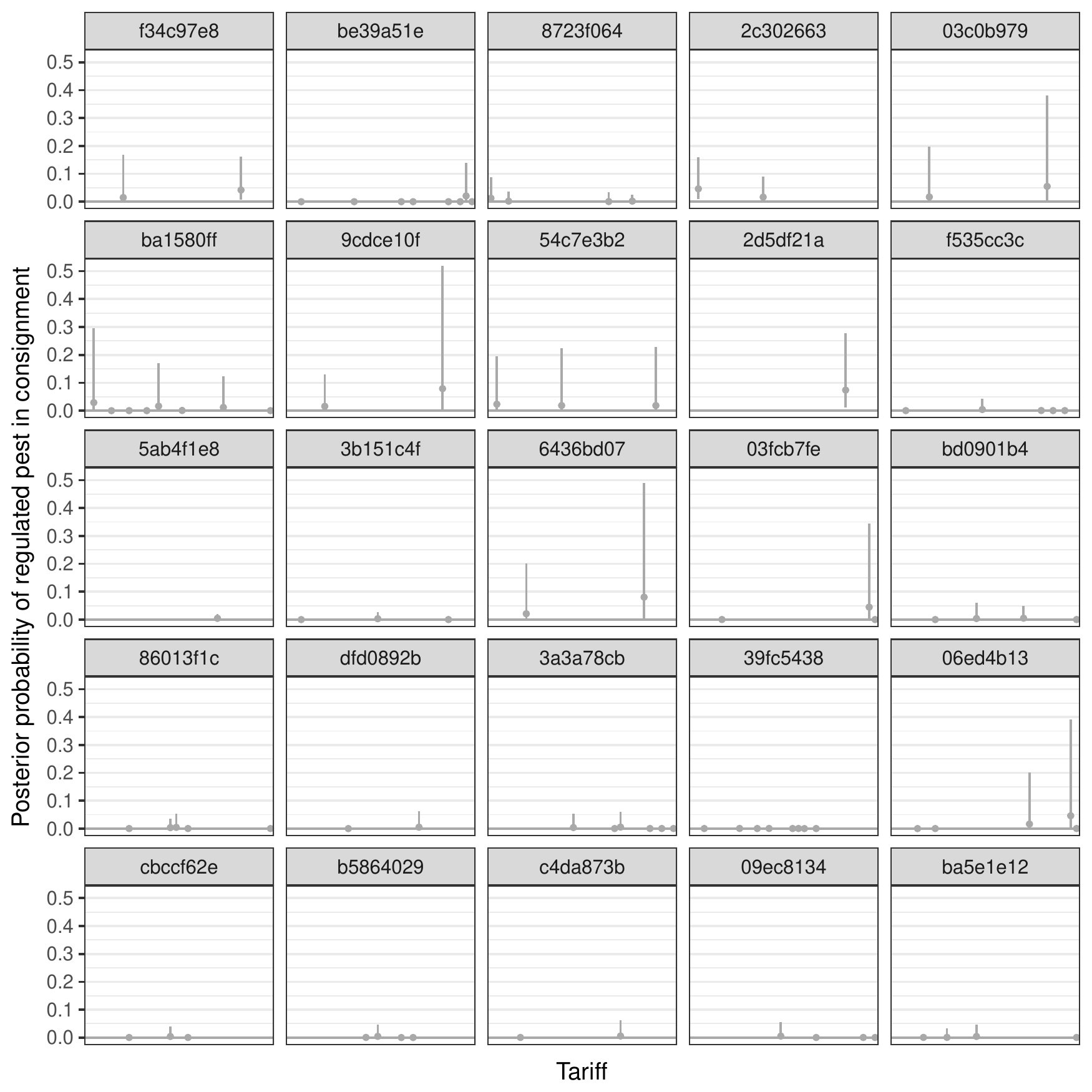} }
\subfloat[Observed proportion of consignments without regulated pests.\label{fig:suppTarfsBot2}]{\includegraphics[width=0.49\linewidth]{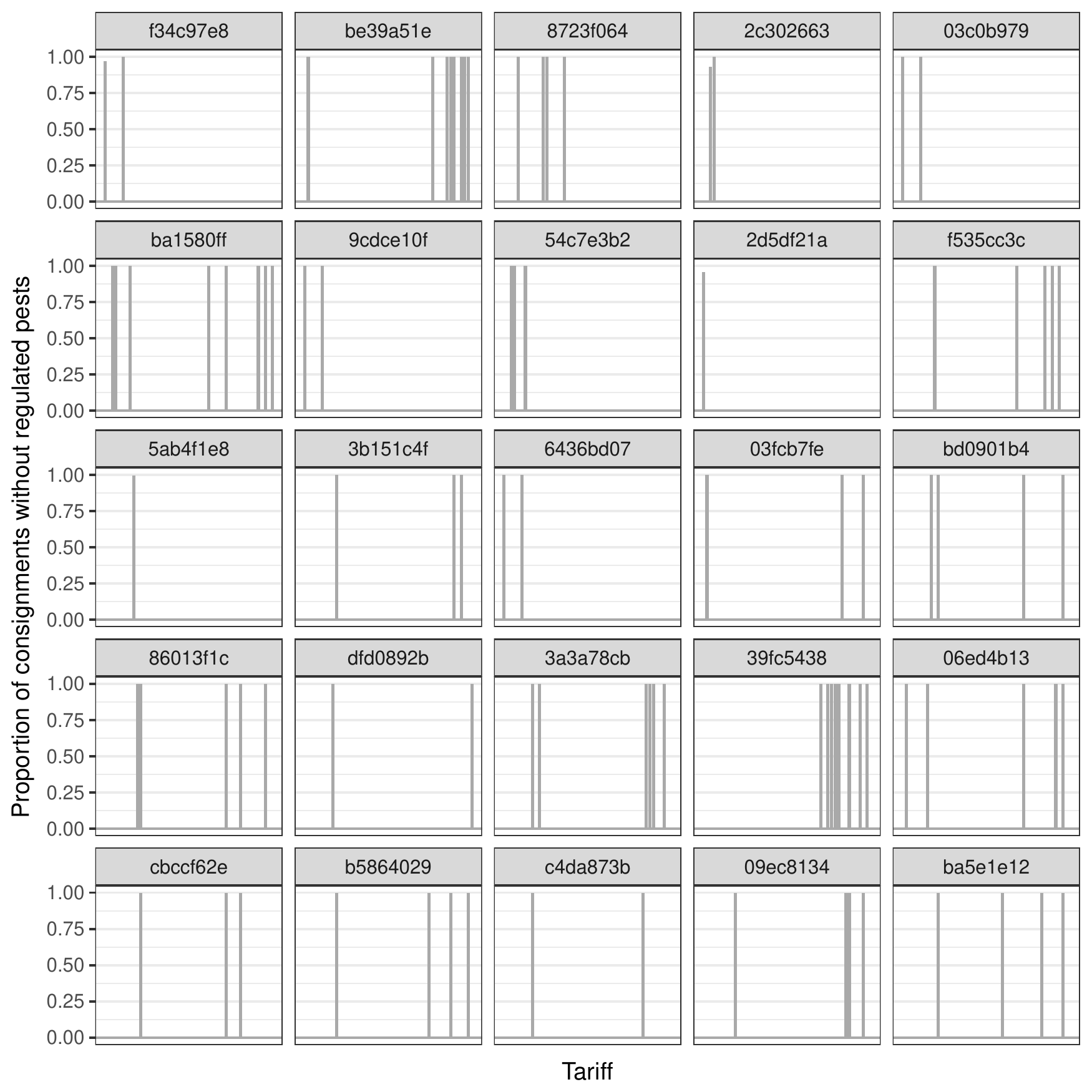} }\caption[Posterior probability of a regulated pest interception in the bottom 25 suppliers, along with the observed proportion of consignments without regulated pests]{Posterior probability of a regulated pest interception in the bottom 25 suppliers, along with the observed proportion of consignments without regulated pests. Panels are ordered left-to-right, top-to-bottom by the probability of their marginal log-odds being less than 0; bars in the left panel show 90\% posterior credible intervals.}\label{fig:suppTarfsBot}
\end{figure}

\end{knitrout}

\section{Discussion and Conclusion}

There was a strong association between regulated pest interceptions and the lower-risk administrative and non-regulated pest interceptions (Section~\ref{sec:assoc-betw-low}). This association was also observed when using administrative interceptions as a predictor for operational profiling (Figure~\ref{fig:roc}), demonstrating the utility of the operational profiling approaches. However we note that this does not carry over into the statistical models (Section~\ref{sec:model-based-sign}), for which including historical rates as predictor variables did not improve model fits.

The statistical profiles still performed well using the cross-validated AUCs (Figure~\ref{fig:mod-aucs}) as well as the predictive log-likelihood. Thus, in answer to our motivating question of which data provide the most useful information about the pathway, we would conclude that it is the knowledge of particular suppliers, tariffs and their combination that is most informative. The previous year's regulated pest profile performed well based on AUC, however adding this to the statistical profiles gave no benefit (Table~\ref{tab:anova}). Furthermore, there is limited scope for investigating why a particular supplier may be problematic. Statistical profiling, in comparison, allows decisions to be based on posterior probabilities. For example, we could calculate a supplier's (marginal) probability of having a regulated pest interception. Intervention could then be planned on either the top ranked suppliers (if funds are limited), or all suppliers that meet a threshold.

A benefit of using a statistical modelling approach to investigate profiling is the added level of interrogation possible from the fitted model. In Section~\ref{sec:model-examination} we demonstrated how we can gain a clearer insight into the governance of this pathway. Firstly, by studying the marginal effects of suppliers and tariffs, we can build up a picture of risk without relying on observed rates, which are noisy due to sampling and process error. We can pinpoint which tariffs and suppliers contribute to excessive risk, essentially by an ordering of the marginal log-odds, and then choose to investigate those that have a posterior probability higher than a pre-defined cutoff set by management.

With a list of potentially risky suppliers, we further demonstrated how a manager could gain information into the governance of those suppliers by investigating the posterior predicted probabilities of regulated pest interceptions (Figure~\ref{fig:suppTarfs}). This information could be used to initially examine why a particular supplier may be having trouble with contaminated consignments, and be used to help improve their processes. Likewise, looking at the less risky importers (Figure~\ref{fig:suppTarfsBot}) may provide information on good process that can be shared to the riskier importers.

To summarise, statistical models constructed using inspection data provide sufficient information for profiling purposes, and can be used to further interrogate the governance of multiple pathways and to help identify the processes underlying poor performance on these pathways.

\section{Acknowledgments}

The authors are grateful to
Lindsay Penrose (ABARES, Australian federal Department of Agriculture and Water Resources) for statistical analysis on an early draft not reported here
and Brendan Woolcott (Plant Division, Australian federal Department of Agriculture and Water Resources) for providing the data for the analysis.

\bibliographystyle{apalike}
\renewcommand{\bibname}{Literature Cited}
\addcontentsline{toc}{chapter}{Literature Cited}
\bibliography{aqis,governance}

\clearpage

\begin{appendices}

  \renewcommand\thetable{\thesection\arabic{table}}
  \renewcommand\thefigure{\thesection\arabic{figure}}

  \setcounter{figure}{0}
  \setcounter{table}{0}

\section{Calculation of rates using empirical Bayes}
\label{sec:calc-rates-using}

In this appendix, we detail the procedure used for calculating the smoothed rates in Section \ref{sec:data} via empirical Bayes. Recall that we have $X_{sty}$ the number of failures out of $n_{sty}$ inspections from tariff $t$ performed in year $y$ from supplier $s$; we assume that $X_{sty} \stackrel{d}{=} \mbox{Binomial}(p_{sty}, n_{sty})$.

To find the empirical Bayes estimate of $p_{sty}$ for supplier $s$, in tariff $t$ and year $y$, assume that the binomial proportions $p_{sty}$ have a prior Beta distribution: $p_{sty} \sim \text{Beta}\left(\alpha_{ty}, \beta_{ty}\right)$. Then $X_{sty}$ has a Beta-binomial distribution, with probability mass function
\begin{align*}
  \Pr\left(X_{sty} = k\right) & = {n_{sty} \choose k} \frac{\Gamma\left(n_{sty} + 1\right)}{\Gamma\left(k + 1\right)\Gamma\left(n_{sty} - k + 1\right)} \frac{\Gamma\left(k + \alpha_{ty}\right)\Gamma\left(n_{sty} - k + \beta_{ty}\right)}{\Gamma\left(n_{sty} + \alpha_{ty} + \beta_{ty}\right)} \frac{\Gamma\left(\alpha_{ty} + \beta_{ty}\right)}{\Gamma\left(\alpha_{ty}\right)\Gamma\left(\beta_{ty}\right)}.
\end{align*}

\noindent The parameters $\alpha_{ty}$ and $\beta_{ty}$ are found using maximum likelihood:
\begin{align*}
  \left(\hat{\alpha}_{ty}, \hat{\beta}_{ty}\right) & = \argmax_{\alpha_{ty}, \beta_{ty}} \left\{{-}\sum_{s = 1}^{S}\log \Pr\left(X_{sty} = x_{sty}\right)\right\}
\end{align*}

\noindent where $x_{sty}$ is the observed value of $X_{sty}$, and $S$ is the number of suppliers. To complete the calculation, the rates for supplier $s$ in tariff $t$ and year $y$, are updated using the following formula:
\begin{align*}
  \tilde{p}_{sty} & = \frac{x_{sty} + \hat{\alpha}_{ty}}{n_{sty} + \hat{\alpha}_{ty} + \hat{\beta}_{ty}}
\end{align*}

\clearpage

\section{Further results}\label{sec:further-results}

\begin{table}[!h]
\caption{Summary of AUC values for profiling strategies, by year.  The profiles are as follows: \textbf{Tariff and Supplier} refers to profiles constructed for the interaction of tariff and supplier, \textbf{Supplier within Tariff} refers to averaging the supplier interception rates within tariffs, and \textbf{Supplier within Tariff, Smoothed} refers to the using the empirical Bayes estimate of the suppliers within tariffs and years. \emph{Regulated pest} refers to using the previous year's regulated pest interception rate; \emph{Non-regulated pest} refers to using the previous year's non-regulated pest interception rate; \emph{Administrative} refers to using the previous year's administrative interception rate; and \emph{Combined} refers to using the previous year's combined interception rate. Each AUC is computed using the data from the \emph{following} year's inspections.\label{tab:aucsAll}} 
\begin{center}
\begin{tabular}{lrrrr}
\toprule
\multicolumn{1}{l}{Profile}&\multicolumn{1}{c}{2008}&\multicolumn{1}{c}{2009}&\multicolumn{1}{c}{2010}&\multicolumn{1}{c}{2011}\tabularnewline
\midrule
{\bfseries Tariff and Supplier}&&&&\tabularnewline
~~Regulated pest&0.881&0.899&0.902&0.917\tabularnewline
~~Non-regulated pest&0.849&0.878&0.859&0.890\tabularnewline
~~Administrative&0.728&0.767&0.759&0.776\tabularnewline
~~Combined&0.833&0.861&0.854&0.867\tabularnewline
\midrule
{\bfseries Supplier within Tariff}&&&&\tabularnewline
~~Regulated pest&0.861&0.897&0.903&0.906\tabularnewline
~~Non-regulated pest&0.839&0.880&0.864&0.907\tabularnewline
~~Administrative&0.795&0.841&0.824&0.842\tabularnewline
~~Combined&0.836&0.881&0.871&0.901\tabularnewline
\midrule
{\bfseries Supplier within Tariff, Smoothed}&&&&\tabularnewline
~~Regulated pest&0.902&0.929&0.935&0.939\tabularnewline
~~Non-regulated pest&0.870&0.909&0.892&0.920\tabularnewline
~~Administrative&0.743&0.815&0.803&0.813\tabularnewline
~~Combined&0.859&0.896&0.890&0.905\tabularnewline
\bottomrule
\end{tabular}\end{center}
\end{table}

\end{appendices}

\end{document}